\documentclass[12pt,leqno]{article}
\usepackage{amsmath,amsthm,amssymb}

\def\init{\setcounter{equation}{0}}
\newtheorem{theorem}{Theorem}[section]
\newcommand{\R}{\mathbb{R}}

\usepackage{tikz}
\usetikzlibrary{decorations.pathreplacing}

\title{A simple approach to temporal cloaking}

\author{G.Eskin, \ \ \  Department of Mathematics, UCLA,\\ Los Angeles,
CA 90095-1555, USA. \ E-mail: eskin@math.ucla.edu
}

\begin{document}
\maketitle

\begin{abstract}
In recent years  a remarkable progress was made in the construction  of spatial cloaks  using the methods of transformation  optics and 
metamaterials.

The temporal cloaking,  i.e. the cloaking of an event in spacetime, was also widely studied by using transformations on spacetime domains.

We propose a simple and general method for the construction  of temporal cloaking using the change of   time variable
only.

\end{abstract}

\section{Introduction}
\init

The transformation optics approach,  combined with the use of metamaterials, leads to a remarkable progress 
in  the construction  of  spatial cloaking devices  and  other  problems  (cf.  J.B. Pendry et al [15],   U. Leonhardt [12] and  others,
see also  [1],  [16]).
The  mathematical analysis of the spatial cloaking  was done by  A. Greenleaf,  Y. Kurylev,  M. Lassas,  and  G. Uhlmann 
  [4],  [5],  [6],  [7]   (see also   [13]).

The temporal cloaking  uses the transformation  of variables  in spacetime. 
The important   works on the temporal cloaking were done  by  physicists  (cf.  McCall,  Kinsler  et al  [8], [10], [11], [14],
 M.  Fridman [3] and others).

In  the above works  mostly the cloaking  in one direction was studied,  i.e.  no complete cloaking in $n\geq 2$  dimensions was achieved.

Our approach to the temporal  cloaking  is based  on a completely new idea   consisting   of  the  change of the time variable 
only.   This way we get a perfect cloaking region.

To preserve  the hyperbolicity of the wave equation
after the change of time variable  one needs to restrict  the size of the cloaked region depending on the wave speed:  
The  size  of the cloaking   region  is  decreasing  when  the speed is increasing.   Thus our approach is better  when  the speed is
not too large as in the case of
 the acoustic equations  and the equations of linear elasticity but  for the optics  problems  the cloaked  region  is too small.

Now we shall briefly describe the content  of the paper.

 In  \S 2
we  apply our method  to the initial boundary value problem for the classical wave equation.  

In \S3   we   give a physical  interpretation   of  the  results  of \S 2.

Then  in \S  4
we extend the results of \S2  to the cases of more general equations.

\section{The main result}
\init

Let  $D$  be a bounded domain in $\R^n$.   Consider  the initial-boundary value problem:
\begin{align}																		\label{eq:2.1}
&Lu (x_0,x)\stackrel{def}{=}\frac{\partial^2 u(x_0,x)}{\partial x_0^2}-
a^2\sum_{k=1}^n\frac{\partial^2 u(x_0,x)}{\partial x_k^2}=0,
\ \ (x_0,x)\in \R\times D,
\\
\label{eq:2.2}
&u=0 \ \ \ \mbox{for}\ \ \ x_0\ll 0,\ x\in D,
\\
\label{eq:2.3}
& u\big|_{\R\times \partial D}=f,
\end{align}
where
$x=(x_1,...,,x_n),\ a$  is a wave speed, $f$ has  a compact support  in $\R\times\partial D$.

Make a change of variables  in $\R\times D$
\begin{equation}																	\label {eq:2.4}
y_0=\varphi_0(x_0,x),\ \ y=x.
\end{equation}

Let  $\tilde L v(y_0,y)=0$  be  the equation  (\ref{eq:2.1})  in $(y_0,y)$-coordinates,  where  $v(y_0,y)=u(x_0,x),\ (y_0,y)$  and  $(x_0,x)$
are related   by  (\ref{eq:2.4}).  
In particular  case  when  
$$
\varphi_0(x_0,x)=x_0+c(x),
$$
we have 
$$
\frac{\partial u}{\partial x_j}=\frac{\partial v}{\partial y_j}  +c_{y_j}(y)\frac{\partial v(y_0,y)}{\partial y_0},\ \ \ 
\frac{\partial u}{\partial x_0}=\frac{\partial v}{\partial y_0}  
$$
and
the equation  $\tilde L v(y_0,y)=0$  has  the  following form
\begin{multline}																	\label{eq:2.5}
\frac{\partial^2 v}{\partial  y_0^2}-a^2\sum_{j=1}^n
\Big(\frac{\partial }{\partial y_j}+c_{y_j}(y)\frac{\partial}{\partial y_0}\Big)
\Big(\frac{\partial}{\partial y_j}+c_{y_j}(y)\frac{\partial }{\partial y_0}\Big)v=0.
\end{multline}
The  symbol of (\ref{eq:2.5})  is  
$$
p(y,\eta)=\eta_0^2-a^2\sum_{j=1}^n(\eta_j+c_{y_j}(y)\eta_0)^2
$$
  or
$$
p(y,\eta)=\Big(1-a^2\sum_{j=1}^n c_{y_j}^2\Big)\eta_0^2-2a^2\sum_{j=1}^n a^2 c_{y_j}\eta_0\eta_j-a^2\sum_{j=1}^n\eta_j^2.
$$
It is strictly hyperbolic   with respect  to $y_0$   (cf.  [9],  \S 23.2,  or  [2], \S 48)  if
\begin{equation}																\label{eq:2.6}
1-a^2\sum_{j=1}^n c_{y_j}^2>0.
\end{equation}
Indeed,  $p(y,\eta_0,\eta)=0$  has   two
 distinct real  roots
\begin{equation}												 				\label{eq:2.7}
\eta_0=\frac{a^2\, c_y\cdot\eta\pm\sqrt{a^4(c_y\cdot \eta)^2+(1-a^2|c_y|^2)a^2|\eta|^2}}{1-a^2|c_y|^2},
\end{equation}
when (\ref{eq:2.6}) holds.  Here  $c_y\cdot\eta=\sum_{j=1}^nc_{y_j}\eta_j$.  Note that when  $n\geq 2$
and $c_y\cdot\eta=0$,  the roots  in (\ref{eq:2.7})  will be not  real  when  $1-a^2|c_y|^2<0$,  i.e.  (\ref{eq:2.5})  is not 
strictly
hyperbolic  when  
(\ref{eq:2.6})  is not satisfied.

Thus the initial  boundary value problems
 are well-posed  when (\ref{eq:2.6})  holds.

We specify that 
\begin{align}																	\label{eq:2.8}
&\varphi_0(x_0,x)=x_0+c(x)\ \ \ \ \mbox{for}\ \ \ x_0\geq 0,
\\ 
&\varphi_0(x_0,x)=x_0\ \ \ \ \mbox{for}\ \ \ x_0<0,                                                        \label{eq:2.9}
\end{align}
where 
\begin{equation}																\label{eq:2.10}
c(x)=c_0\chi(x),\ \  c_0>0,
\end{equation}
$\chi(x)\in C_0^\infty(\R^n),\chi(x)=0$  for  
$|x|>c_1,\chi(x)=1$  for  $|x|<\frac{c_1}{2},0<\chi(x)<1$  for  $\frac{c_1}{2}<|x|<c_1$, the ball $|x|\leq  c_1$  is inside $D$. 
Therefore  $y_0=\varphi_0(x_0,x)$  is  strictly  increasing  in  $x_0\in \R$   and  has a jump  $\varphi_0(+0,x)-\varphi_0(-0,x)=c(x)$  at  $x_0=0$.


Denote   by   $Y^+$  and  $Y^-$  the sets
$\{y_0 \geq c(y),y\in \R^n\}$  and  $\{y_0 < 0,y\in \R^n\}$,  respectively.

Let

\begin{equation}																\label{eq:2.11}
Y_0=\{(y_0,y):0<y_0<c(y), |y|<c_1\}.
\end{equation}
Note that
\begin{equation}																\label{eq:2.12}
Y^+\cup \overline Y^-=\R^{n+1}\setminus Y_0,
\end{equation}
and $y_0=\varphi(x_0,x),y=x$  maps  $\R\times\R^n$  onto  $Y^-\cup Y^+$.

Consider the equation
\begin{equation} 																\label{eq:2.13}
\tilde L^-v^-(y_0,y)=0\ \ \ \mbox{in}\ \ \ Y^-\cap(\R\times D),
\end{equation}
where$\tilde L^-=L,\ y_0=x_0<0$  
with the initial condition 
\begin{equation} 																\label{eq:2.14}
v^-=0\ \ \   \mbox{for}\ \ \   y_0\ll 0,y\in D,
\end{equation}
and the boundary condition 
\begin{equation}																\label{eq:2.15}
v^-=f\ \ \ \mbox{on}\ \ \ (\R\times\partial D)\cap Y^-.
\end{equation}
This problem is well-posed and has a unique smooth solution  $v^-(y_0,y)$.  Now consider  the initial-boundary  value  problem  
for
\begin{equation}																\label{eq:2.16}
\tilde L^+v^+(y_0,y)=0\ \ \ \mbox{in}\ \ \ Y^+\cap(\R\times D),
\end{equation}
corresponding to the change  of variable  $y_0=\varphi_0(x_0,x)=x_0+c(x),\ x_0>0,$  with the boundary  condition		 
\begin{equation}																\label{eq:2.17}
v^+=f\ \ \ \mbox{on}\ \ \ (\R\times\partial D)\cap Y^+,
\end{equation}
and the initial conditions
\begin{equation}																\label{eq:2.18}
v^+\Big|_{\partial Y^+\cap (\R\times D)}=v^-\Big|_{\partial Y^-\cap (\R\times D)},\ \ \ 
\frac{\partial v^+}{\partial y_0}\Big|_{\partial Y^+\cap (\R\times D)}
=\frac{\partial v^-}{\partial y_0}\Big|_{\partial Y^-\cap (\R\times D)}.
\end{equation}
Note that we assume that $v^-\big|_{\partial Y^-\cap D}$  and $\frac{\partial v^-}{\partial y_0}\big|_{\partial Y^-\cap D}$  are  already known.

 This initial-boundary value problem also has a unique  smooth  solution  when  (\ref{eq:2.6})  holds.  
 Therefore  we can determine a function  $v(y_0,y)$  such 
 that $v=v^+$  in  $Y^+\cap(\R\times D),\ v=v^-$  in  $Y^-\cap (\R\times D)$,   $\tilde L^+v^+=0$  in  $Y^+\cap\R\times  D,\ \tilde L^-v^-=0$  in  
 $Y^-\cap(\R\times D),\ \ v=v^-=0$  for  $y_0\ll 0,\ y\in D,\ v\Big|_{\R\times\partial D}=f$  and $v^+$  and  $v^-$  satisfy  the conditions 
  (\ref{eq:2.18}).

Once $v(y_0,y)$  is given  we determine  $u^+(x_0,x)$   for  $x_0>0, $ such  that   $u^+(x_0,x)=v^+(y_0,y)$,
  where 
$x_0=y_0-c(y),x=y, (y_0,y)\in  Y^+$.
Analogously,  $u^-(x_0,x)=v^-(y_0,y)$,  where  $x_0=y_0, x=y, (y_0,y)\in Y^-$.

Then  $u^+(x_0,x)$   and  $u^-(x_0,x)$  satisfy  the wave  equation  (\ref{eq:2.1})  for $x_0>0$  and  $x_0 <0$,  respectively.
Since the conditions  (\ref{eq:2.18})  are satisfied,  we have  that
\begin{align}																	\label{eq:2.19}
&\lim_{\overset{x\rightarrow 0}{x_0<0}} u^-(x_0,x)=\lim_{\overset{x_0\rightarrow 0}{x_0>0}} u^+(x_0,x)
\\
\nonumber
&\lim_{\overset{x\rightarrow 0}{x_0<0}} \frac{\partial u^-(x_0,x)}{\partial x_0}=\lim_{\overset{x\rightarrow 0}{x_0>0}} \frac{\partial u^+}{\partial x_0}(x_0,x)
\end{align}
Therefore $u(x_0,x)=u^+(x_0,x)$  for  $x_0>0$  and  $u(x_0,x)=u^-(x_0,x)$  for  $x_0<0$  satisfied (\ref{eq:2.1})
 in  $\R\times D$ and also
satisfies   the initial and boundary conditions (\ref{eq:2.2}),  (\ref{eq:2.3}).

Since  $c(x)=0$  in  $D\setminus\overline B$,  where  $B=\{x: |x| < c_1\}$  we have that  $x_0=y_0, x=y$  in $\R\times(D\setminus \overline B)$.
Thus
\begin{equation}																\label{eq:2.20}
u(x_0,x)=v(x_0,x)\ \ \ \ \mbox{in}\ \ \R\times (D\setminus \overline B).
\end{equation}
Therefore  the boundary data of $u(x_0,x)$  and  $v(y_0,y)$  on $\R\times \partial D$  are  the same.  This implies  that one can 
not distinguish  between  $u(x_0,x)$  in  $\R\times D$  and   $v(y_0,y)$  in  $(\R\times D)\cap(Y^+\cup Y^-)$
using the boundary measurements. 
Since the domain $Y_0$  is outside of $Y^+\cup \overline Y^-,\ Y_0$  is a temporal 
cloaking domain  and the observer  on 
$\R\times \partial D$  does not suspect  its existence.
\
\\
{\bf Remark 2.1}
\ Note that the change of variables  (\ref{eq:2.6})  is not singular as it happens in the case of a  spatial
cloaking domain.

Summarizing the results   of  this section  we get  the following theorem:
\begin{theorem}                             											\label{theo:2.1}
Let domains  $Y_0,Y^+,Y^-$  be  the same  as in (\ref{eq:2.11}),  (\ref{eq:2.12})  and  let operators $\tilde L^-$  and  $\tilde L^+$  be
the same  as in  (\ref{eq:2.13})  and  (\ref{eq:2.16}).  
Let the condition (\ref{eq:2.6})  holds.
Consider  the solution  $v^-(y_0,y)$  
of the initial-boundary value  problem (\ref{eq:2.13}),  (\ref{eq:2.14}),  (\ref{eq:2.15})  in  $Y^-\cap (\R\times D)$.  


Let  $v^+(y_0,y)$  be the solution  in $Y^+\cap(\R\times D)$  of  the  initial-boundary  value problem  
(\ref{eq:2.16}),  (\ref{eq:2.17}),  (\ref{eq:2.18}).  

Let  $v(y_0,y)=v^-(y_0,y)$  for $y_0\leq 0,\ v(y_0,y)=v^+(y_0,y)$  for  $y_0\geq c(y)$,  i.e.  $v(y_0,y)$  is the solution  in  
$\overline Y^+\cup Y^-=(\R^n\times D)\setminus Y_0$.

Let  $u^+(x_0,x)=v^+(y_0,y)$  for  $x_0>0$,  where   $x=y, x_0=y_0-c(y)$   and let  $u^-(x_0,x)=v^-(y_0,y)$    for  $x_0<0$  where  
$x=y, x_0=y_0$.  It follows from  (\ref{eq:2.18})   that  $u(x_0,x)=u^+(x_0,x)$  for  $x_0>0,  u(x_0,x)=u^-(x_0,x)$  for  $x_0<0$
extends to a smooth  function  in $\R\times D$  that  satisfies (\ref{eq:2.1}),  (\ref{eq:2.2}),  (\ref{eq:2.3}).   
The boundary measurements  of  $v(y_0,y)$  defined in  $(\R\times D)\setminus Y_0$  and  $u(x_0,x)$  defined  in  $\R\times D$  are  equal. 

Thus $Y_0$  is a perfect  temporal  cloak.
\end{theorem}

\begin{tikzpicture}
[scale=2]
\draw[->](0,-.4)-- (0,1.2);
\draw[->](-2.5,0)-- (2.5,0);
\draw[ultra thick] (-2,0)-- (2,0);

\draw[ultra thick](-2,0).. controls (-1.6,0.4)  and (-1,0.8) .. (0,0.85);
\draw[ultra thick](0,0.85).. controls (1,0.8)  and (1.6,0.4) .. (2,0);

\draw (0.3,0.5) node {$Y_0$};
\draw (1,1.) node {$Y^+$};
\draw (1,-0.2) node {$Y^-$};

\draw (0.2,1.3) node {$y_0$};
\draw (2.7,0.1) node {$y$};

\end{tikzpicture}
\\
\\
\
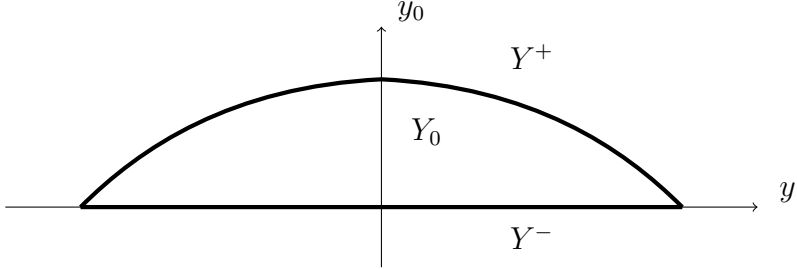
{\bf Fig. 1.} Domain $Y_0$  is  a temporal cloaked region. 

\section{Physical interpretation  of  Theorem \ref{theo:2.1}}
\init

Assume,  for the  definiteness,  that (\ref{eq:2.1})
describes  the vibration  of a membrane where   $a=\sqrt{\frac{T}{\rho}}$   is  the speed,   $T$  is the tension  and   $\rho$   is the density.   Let  
$v^-(y_0,y)$     be the  solution  of (\ref{eq:2.1}),  (\ref{eq:2.2}),   (\ref{eq:2.3})   for  $y_0\leq 0$.  We  shall call  $V^-(y_0,y)$   the physical solution  of
(\ref{eq:2.1})  (to distinguish  from the numerical  solution   $v^-(y_0,y)$),  i.e.  $V^-(y_0,y)$  is
the actual   vibrating  membrane  where  $V^-(y_0,y)$   is  the position of the membrane  at point  $y$  and  at time   $y_0\leq 0$.

Consider  now   the equation  $\tilde L^+v^+(y_0,y)=0$  in $Y^+$   with the boundary  condition  (\ref{eq:2.3})
and  the initial  conditions  (\ref{eq:2.18}).   Denote by $w(x_0,x_1)$  the  function  $v^+(y_0-c(y),y)$  where  $x_0=y_0-c(y)\geq 0,  y=x$.
Function $w(x_0,x)$  is  the  solution  of   (\ref{eq:2.1})   for   $x_0\in  [0,+\infty)$  with   the  boundary condition  (\ref{eq:2.3})  and the  initial   
conditions
\begin{equation}			       														\label{eq:3.1}
w(0,x)=v^-(0,x),\ \ \frac{\partial w(0,x)}{\partial  x_0}=\frac{\partial v^-(0,x)}{\partial x_0},
\end{equation}
(cf.  (\ref{eq:2.18})).  Knowing  $w(x_0,x)$  we can  recover   $v^+(y_0,y)$   by the formula
\begin{equation}																	\label{eq:3.2}
v^+(y_0,y)=w(y_0-c(y),y),\ \ \ (y_0,y)\in  Y^+.
\end{equation}
Let  $W(x_0,x)$   be  the physical  solution  of  (\ref{eq:2.1})  for $x_0>0$,  i.e.  the actual   membrane  vibrationg  on  $(0,+\infty)$.

It follows   from  (\ref{eq:3.2})    that  $V^+(y_0,y)=W(y_0-c(y),y)$  is the physical  solution  of  (\ref{eq:2.5})  on  $Y^+$,   i.e.   $V^+(y_0,y)$  is 
the actual   vibrating membrane  shifted in time.

It follows   from  (\ref{eq:2.1})   that  the physical initial data  of  $V^+(y_0,y)$    on  $y_0-c(y)=0$ are equal  to the physical final  data   of  
$V^-(y_0,y)$  on  $y_0=0$,  i.e.
\begin{equation}																	\label{eq:3.4}
V^+(y_0,y)\Big|_{y_0-c(y)=0}=V^-(0,y),\ \ \ \ \frac{\partial V^+}{\partial y_0}\Big|_{y_0-c(y)=0}=\frac{\partial V^-(0,y)}{\partial  y_0},
\end{equation}
since $W(0,y)=V^-(0,y),\ \frac{\partial W(0,y)}{\partial y_0}=\frac{\partial V^-(0,y)}{\partial  y_0}$.

Summarizing,   we have  a physical solution $V^-(y_0,y)$   (membrane)   on  $Y^-$   satisfying  initial condition  (\ref{eq:2.2})   and boundary 
condition  (\ref{eq:2.3}),   and we have another physical   solution  $V^+(y_0,y)$  on  $Y^+$  satisfying the boundary condition  (\ref{eq:2.3}).  
The initial physical  condition  of  $V^+(y_0,y)$  are  equal  to the final  physical   condition  of  $V^-(y_0,y)$.   Note that
the cloaking region $Y_0$  is outside of domains  $Y^-$   and  $Y^+$  of these two physical  solutions  $V^-$  and  $V^+$.
If one takes the physical    measurement  of the force  $T\frac{\partial V^+}{\partial n}$  on the    
boundary  one will get the same result  as when  we measure the force  on the boundary  for the initial  boundary value problem (\ref{eq:2.1}),
(\ref{eq:2.2}),   (\ref{eq:2.3}),    i.e. it is impossible to find out whether  the cloaking region exists.   Here   $\frac{\partial}{\partial n}$  is the normal
 derivative to  $\partial D$. 

\section{More general equations}
\init
Results of \S2 can be easily  extended  to the case of more general  equations.

Consider  a strictly hyperbolic equation in  $\R\times D$  of the form
\begin{equation}																\label{eq:4.1}
Lu\stackrel{def}{=}\sum_{j,k=0}^n\frac{1}{\sqrt{(-1)^ng(x_0,x)}   }\,
\frac{\partial}{\partial x_j}
\Bigg(\sqrt{(-1)^ng(x_0,x)}\ \,g^{jk}(x_0,x)\frac{\partial u(x_0,x)}{\partial x_k}\Bigg)=0,
\end{equation}
where  $g^{-1}(x_0,x)=\det [g^{jk}(x_0,x)]_{j,k=0}^n,\ g^{00}(x,t)>0,\ \det [g^{jk}(x_0,x)]_{j,k=1}^n\neq 0$.

We assume that the initial and boundary conditions (\ref{eq:2.2}),  (\ref{eq:2.3})  are satisfied.  Note that we allow  the metric to be 
time-dependent.

We assume also that the boundary  $\R\times \partial D$  is  time-like,  i.e.
\begin{equation}																\label{eq:4.2}
\sum_{j,k=0}^ng^{jk}(x_0,x)\nu_j\nu_k<0\ \ \ \mbox{on}\ \  \R\times \partial D,
\end{equation}
where $(\nu_0,\nu_1,...,\nu_n)$  is the normal  to $\R\times \partial D$.
Then the initial-boundary value problem  (\ref{eq:4.1}),  (\ref{eq:2.2}),  (\ref{eq:2.3}) is
well-posed  (cf.  [9],  \S 23.2).

Consider the change of variables
\begin{equation}																	\label{eq:4.3}
y_0=\varphi_0(x_0,x),\ \ \ y_k=x_k,\  1\leq k\leq n.
\end{equation}

The equation  (\ref{eq:4.1})
has the following form in  $(y_0,y_n)$ coordinates  (cf.  (\ref{eq:2.5})):   
\begin{equation}																\label{eq:4.4}
\hat Lv\stackrel{def}{=}\sum_{j,k=0}^n\frac{1}{\sqrt{(-1)^n \hat g(y_0,y)}   }\,
\frac{\partial}{\partial y_j}
\Bigg(\sqrt{(-1)^n \hat g(y_0,y)}\ \,\hat g^{jk}(y_0,y)\frac{\partial  v(y_0,y)}{\partial y_k}\Bigg)=0,
\end{equation}
where
\begin{align}																	\label{eq:4.5}
&\hat g^{jk}(y_0,y)=g^{jk}(x_0,x),\ \ \ \ 1\leq j,k\leq n,
\\																				\label{eq:4.6}
&\hat g^{00}(y_0,y)=\sum_{p,r=0}^n g^{pr}(x_0,x)\frac{\partial \varphi_0}{\partial x_p}\frac{\partial \varphi_0}{\partial x_r},
\\																				\label{eq:4.7}
&\hat g^{0j}(y_0,y)=\sum_{p=0}^n g^{pj}(x_0,x)\frac{\partial \varphi_0}{\partial x_p}.
\end{align}
We consider the case when  $\varphi_0(x_0,x)$  is 
arbitrary   strictly  increasing in $x_0$  piece-wise smooth function having a jump  at  $x_0=0$.

We get from  $u(x_0,x)=v(y_0,y)$  as  in  \S 2
\begin{equation}																	\label{eq:4.8}
u_{x_j}=v_{y_j}+\varphi_{0x_j}v_{y_0}, \ 1\leq j\leq n,\ \ u_{x_0}=\varphi_{0x_0}v_{y_0}.
\end{equation}
Therefore equation  (\ref{eq:4.4})  can be written  in the form  similar  to (\ref{eq:2.5}).

Let  $x_0=\psi(y_0,y)$  be the  inverse  to  $\varphi_0(x_0,y)$,  i.e.   $\varphi_0(\psi(y_0,y),y)=y_0$.
As  in  (\ref{eq:2.6})  the equation  (\ref{eq:4.4})  will be hyperbolic  with respect to  $y_0$   if 
\linebreak
 (cf.  [9],  \S23.2)
\begin{equation}																	\label{eq:4.9}
\tilde g^{00}(y_0,y_0)=\sum_{p,r=0}^n g^{pr}\frac{\partial\varphi_0}{\partial x_r}\frac{\partial \varphi_0}{\partial x_p}>0.
\end{equation}
Note that  (\ref{eq:4.9})  coincides with  (\ref{eq:2.6})  when  (\ref{eq:4.1})  has the form  (\ref{eq:2.1}).

To simplify  the condition  (\ref{eq:4.9})   consider  a particular   case  when  $g^{00}>0, g^{0j}=g^{j0}=0,  \ 1\leq j\leq n$,  and
\begin{equation} 																\label{eq:4.10}
\sum_{j,k=1}^n  g^{jk}\xi_j\xi_k\leq -C_0(\xi_1^2+\xi_2^2+\xi_n^2),
\end{equation}
i.e. the case   when  the spatial  part  of the equation  (\ref{eq:4.1})  is elliptic.  Let $\varphi_0=y_o-c(y)$  as in \S2.  Then the inequality (\ref{eq:4.9})
has the form
\begin{equation}										 						\label{eq:4.11}
g^{00}+\sum_{j,k=0}^n g^{jk}c_{y_j}c_{y_k}>0.
\end{equation}
  Thus 
 we get from  (\ref{eq:4.10})  and  (\ref{eq:4.11})
that  $|c_y|^2\leq  \frac{1}{C_0}g^{00}$.

  For  the well-posedness  of the initial-boundary value problem  the  boundary
  $\R\times \partial D$  must  be time-like,   i.e.
  \begin{equation}																\label{eq:4.12}
  \sum_{j,k=0}^n\hat g  ^{jk}(y)\nu_j\nu_k<0 \ \ \ \mbox{on}\ \ \ \R\times \partial D.
  \end{equation}
Note that (\ref{eq:4.12})  follows  from  (\ref{eq:4.2})  since  $\nu_0=0$  and  $\hat g^{jk}=g^{jk}$  for  $1\leq j,k\leq n$.
Therefore  Theorem \ref{theo:2.1}   holds  also for the equation  (\ref{eq:4.1}),  assuming that (\ref{eq:4.9})  holds.

Similar   results hold  for  the hyperbolic systems, for  example,  for the elasticity equations.



\begin{thebibliography}{9999}
\bibitem[1]{} T. Ergin, N. Stenger, P. Brenner,  J. Pendry,  M.  Wegener,  Science  328  337-339 (2010)
\bibitem[2]{} G.Eskin,  Lectures  on linear  partial  differentialequations,  AMS,  GSM  vol. 123  (2011)
\bibitem[3]{} M. Fridman et al,  Demonstration  of temporal cloaking,  Nature,  481  (2012),  62.
\bibitem[4]{} A. Greenleaf, M. Lassas  and G. Uhlmann,  On nonuniqueness for Calderon's inverse  problem,
Math. Res. Lett.,  10 (2003), 685-693
\bibitem[5]{} A. Greenleaf,  Y. Kurylev,  M. Lassas and  G.  Uhlmann,  Full-wave  invisibility  of active devices  at all
frequencies,  Comm. Math. Phys., 275 (2007),  749-789
 \bibitem[6]{} A. Greenleaf,  Y. Kurylev,  M. Lassas and  G.  Uhlmann,   Invisibility and inverse problems,  Bull. Amer. Math. Soc.,  46 (2009),  55-97 
\bibitem[7]{} A. Greenleaf,  Y. Kurylev,  M. Lassas and  G.  Uhlmann,  Cloaking devices,  electromagnetic wormholes 
and transformation  optics,  SIAM Review,  51  (2009),  3-33
\bibitem[8]{} J. Gratus,  P.  Kinsler,  M. McCall,  R. Thompson,  On spacetime transformation  Optics  in Temporal  and  Spatial 
Dispersion
\bibitem[9]{} L. Hormander, The Analysis of Linear Partial
Differential Operators, vol. 3, Chapter 24,   Springer Verlag  (1985)
\bibitem[10]{} P. Kinsler,  M. McCall,   Cloaks, editors and bubbles:  application  of spacetime  transformation theory,
Ann. Phys.  (Berlin) 526, No 1-2, 51-62 (2014)
\bibitem[11]{} P. Kinsler,  M. McCall,   Generalized Transformation Design:  metrics,  speeds  and  diffusion,ArXiv:1510.06890
\bibitem[12]{} U. Leonhardt,  Optimal  conformal  mapping,  Science, 312 (2006) 1777-1780
\bibitem[13]{}  H.Liu and  G.Uhlmann,  Regularized  transformation  optics  cloaking in acoustic  and electromagnetic  scattering,
Inverse problems  and imaging,  111-136,  Panor Syntheses, 44,  Soc. Math. France,  Paris 2015
\bibitem[14]{} M.McCall et al,  A spacetime  cloak,  or a history editor,  Journal of Optics 13  (2011),  024003
\bibitem[15]{} J.B. Pendry, D. Schurig and D.R. Smith,  Controlling electromagnetic fields,  Science, 312 (2006),
1780-1782 
\bibitem[16]{} D. Schurig,  J. Mock,  B. Justice,  S. Cummer,  J. Pendry,  A. Starr,  D. Smith,  2006  Science 314,  977-980  



\end{thebibliography}
\end{document}